\documentclass[11pt,epsfig]{article}
\usepackage{graphicx}

 1
\font\elevenrm=cmr10 scaled\magstep 1

\renewenvironment{thebibliography}[1]
 { \elevenrm
   \begin{list}{\arabic{enumi}.}
    {\usecounter{enumi}     \setlength{\parsep}{0pt}
     \setlength{\itemsep}{3pt} \settowidth{\labelwidth}{#1.}
     \sloppy
    }}{\end{list}}
\begin{document}

\title{{\bf Charmed and beautiful baryonic systems (hypernuclei) 
in chiral soliton models}   \\}
\author{V.B.~Kopeliovich$^{a,b}$\footnote{{\bf e-mail}: kopelio@inr.ru} and
  D.E.~Lanskoy$^c$\footnote{{\bf e-mail}: lanskoy@sinp.msu.ru}
\\
\small{\em a) Institute for Nuclear Research of RAS, Moscow 117312, Russia}
\\
\small{\em b) Moscow Institute of Physics and Technology (MIPT), Dolgoprudny, Moscow district, Russia} 
\\
\small{\em c) Department of Physics, Moscow State University, Moscow, Russia }
}

\date{}
\maketitle

\begin{abstract}
The binding energies of baryonic systems with quantum number charm or beauty and neutron excess
(or large isospin) are estimated within the bound state version of the chiral topological soliton model.
 The procedure of the skyrmion rescaling is applied which is of importance for large enough
flavor excitation energies (for heavy flavors). Inclusion of the flavor excitation energies into
the minimization of the total mass of the baryonic state naturally leads to its decrease, and this
effect increases with increasing flavor mass. Our results concerning the binding energies
turned out to be partly negative.
\end{abstract}

\section{Introduction} 
The nature of baryonic states (nuclear fragments) with very unusual properties,
in particular, with heavy flavors - charm and/or beauty, is of interest not only for the nuclear physics itself, but in cosmology 
and astrophysics as well, because production and subsequent decay
of such states could play an important role at the early stages of the Universe evolution,  which are not well 
understood so far.  

Studies of charmed nuclei are the unique tool to understand the low-energy
interactions of charmed particles. Generally speaking, the Nature
arrangement is based on 6 quark flavors obeying unitary SU(6)
symmetry, though strongly violated. On the other hand, our
knowledge of baryonic interactions is drastically asymmetric in
this sense and mainly corresponds only to the SU(2) subgroup
(nucleons). Strange hypernuclei are known to give the
unique possibility of extension our knowledge to the strange
sector (for reviews, see \cite{HypRev} and other references). Then, studies of charmed and beautiful nuclei is the way to extend the picture of
the low-energy hadronic interactions to the SU(4) and the SU(5) world.
The earlier theoretical treatment of the heavy flavored hypernuclei properties can be found
in \cite{tyap} - \cite{c-triton} mostly within various potential approaches.

The chiral (topological) soliton models provide conceptionally different approach to this problem
which has both some advantages and certain disadvantages. The attempt to estimate  the total binding energies
of heavy flavored hypernuclei within the chiral soliton approach (CSA) was made in \cite{kop-1} where the total binding energies of states with
flavor up to $|F|=2$ have been roughly estimated for baryon numbers up to $B=4$, and somewhat later in \cite{kopzak}
where estimates have been extended up to $B=8$ using the rational map ansatz by Houghton, Manton and Sutcliffe \cite{hms}.
The tendency of some increase of the binding energy with increasing mass of the flavor and baryon number has been
noted in \cite{kop-1,kopzak}. 

Here we study the influence of the change of the scale of the whole skyrmion on the energy of the quantized states with 
heavy flavors. The squeezing of the skyrmion leads to considerable decrease of the energy (mass) of the quantized state,
which is especially important for the charm or beauty quantum numbers. In several cases we obtained the decrease of
the binding energies.

In the next section the historical background of the heavy flavored nuclei studies is overviewed.
In section 3 we describe briefly the chiral soliton model and the quantization procedure.
The static characteristics of multiskyrmions, necessary for the quantization procedure and the spectrum of states
calculations are presented here. Section 4 contains description of the squeezing procedure which leads to considerable 
decrease of the energy of quantized states. Other states are evident enough.

\section{Historical background}
A possibility of existence of the $\Lambda _c$ nuclei was first
discussed in \cite{tyap} soon after the discovery of charm.
Then a number of theoretical papers appeared \cite{Dk}-\cite{Ya} studying properties of charmed
nuclei. Mostly, meson-exchange $\Lambda _cN$ interactions or schematic approaches adopting some model 
$\Lambda _cA$ potentials were utilized. This stage of investigations was summarized in
\cite{Star}. Then the theoretical activity becomes poor, probably
due to lack of empirical information. In the more recent years, few
studies were published \cite{Tsush,Tan,Julia,c-triton}.

So far, charmed nuclei have not been observed confidently. Some candidate events were proposed in 
the emulsion experiment \cite{Exp}. On the other hand, negative results were reported by three other experimental 
groups \cite{Exp2,Exp3,Exp4}, which used nuclear emulsions too. Other ways to produce charmed and bottomed nuclei were discussed 
in \cite{Star,Bres,Fel}.

Theoretical studies of $\Lambda _c$ nuclei naturally use the rich experience achieved in the field of strange hypernuclei. On the other hand, 
$\Lambda _c$-nucleus dynamics differs from the $\Lambda _s$ one in some substantial points. Below, we briefly
discuss main qualitative features of charmed
hypernuclei, following Ref. \cite{Star} to some extent,
emphasizing mostly properties, which are different from those of
strange hypernuclei.
Of course, quantitative predictions unlikely can be reliable in this stage. 

\begin{enumerate}
{\item Seemingly all theoretical considerations predict attraction
between $\Lambda _c$ and nuclei. The intensity of this attraction
varies, however, in
different approaches. When coupling constants $\Lambda _s\Lambda _sm$ and $%
\Lambda _c\Lambda _cm$ for nonstrange (and noncharmed) mesons $m$
(or nonrelativistic $\Lambda _sN$ and $\Lambda _cN$ potentials)
are supposed to be equal to each other (e.g., \cite{Dk,Kol}),
$\Lambda _c$ in light systems appears to be bound deeper by the strong interaction
than $\Lambda _s$ due to the greater mass and corresponding reduction
of the kinetic energy (but see discussion of the Coulomb potential
below). On the other hand, it is possible that the $\Lambda _cN$
interaction is weaker than the $\Lambda _sN$ one (\cite{Star} and
references therein). In this case, the nuclear well for 
  
$\Lambda _c$ is shallower and its depth can be, for instance, only a half
of that for $\Lambda _s$. In the first case (comparable $\Lambda_s N$ and $\Lambda_c N$ interaction), the charmed deuterons are
probably unbound, but the charmed tritons can exist \cite{Dk,Gib,c-triton}.
In the second case (a shallow well) charmed nuclei become bound
only at relatively large $A$ (maybe, at $A>10 $ \cite{Star,Tan}).}

{\item The important role in the $\Lambda _sN$ interaction is
played by kaonic exchange, which leads to Majorana
(space-exchange) interaction. This implies the strong orbital moment
dependence of the interaction: The considerable attraction in the $s$ wave
and relatively weak interaction (maybe even slight repulsion) in
the $p$ wave \cite{LY}. For $\Lambda _c$ hyperon, $D$ meson exchange is
extremely short-range and probably negligible. So one may expect
comparable $s$ and $p$ wave attractions \cite{BN,Ya}. The
situation is similar to that for the double-strange $\Xi $ hypernuclei, where the
well depth probably grows substantially with $A$ \cite{Yam}. Probably, the nuclear potential for $\Lambda _c$ also deepens with $A$. Some
dependence of the well depth on $A$ is seen in \cite{BN}, but this
dependence was not studied so far.}

{\item The important role of Coulomb $\Lambda^+_cA$ repulsion was
shown first in \cite{Kol}. It appears that, contrary to strange
hypernuclei, $\Lambda_c$ binding energy $B_{\Lambda
_c}$ does not saturate, but rather reaches a maximum in the vicinity of Ca (%
$A\sim 40$) and then falls. In the case of a deep well, charmed
nuclei remain bound up to heaviest ones. But for a shallow
well, they exist only in the finite range of $A$ (island, in terms
of \cite{Star}). Even for a relatively deep well, 20 MeV, bound
states exist only at $A<100$ \cite{Tan}. However, the $A$
dependence was studied only with constant well depths. If the well
deepens with $A$, it can provide a competitive effect.}

{\item In the early stage of strange hypernuclear physics, production and decay 
of $\Lambda_s$ hypernuclei were investigated in emulsion experiments \cite{DP}. 
Identification of hypernuclei was performed by their weak decay. 
A hypernucleus can be produced in the ground or some excited state. 
But a discrete excited state deexcites by $\gamma$ emission (nonobserved in emulsions) well before the weak decay. 
Therefore, only hypernuclear ground states were available for study. Kolesnikov et al. pointed out \cite{Kol} that discrete excited states of
$\Lambda _c$ hypernuclei probably decay weekly rather than
electromagnetically since the $\Lambda_c$ lifetime is by few orders of magnitude smaller than that of $\Lambda_s$. 
So, one may expect that not only ground state
can be observed in an emulsion experiment. To our knowledge,
nobody calculated so far rates of electromagnetic transitions
between $\Lambda _c$ hypernuclear levels (they depend strongly on
unknown excitation energies), but evidently the
electromagnetic lifetimes are greater than the $\Lambda _c$ lifetime (2$%
\cdot 10^{-13}$ s) often, if not always. }

%
%
%
%
%

{\item As for heavier charmed hyperons, $\Sigma _c$ can convert ($\Sigma _cN\to\Lambda _cN$)
quickly in a nucleus similarly to $\Sigma _s$.
However, $\Xi _c$ ($C=+1$, $S=-1$, $usc$ or $dsc$) hypernuclei can
be interesting. The key point \cite{Star} is that the energy
release from the conversion $\Xi _c+N\rightarrow \Lambda
_c+\Lambda $ is very small, namely, 5--6 MeV \cite{PDG}. It is possible even that $%
\Xi _c$ hypernuclei are stable with respect to the strong
interaction ($\Xi _c$ cannot convert) due to nuclear binding. Even if this is not the
case, the widths are expectedly rather small. Strange charmed nuclei have
been considered only very briefly \cite {Dk,Star} without taking into account $\Xi _cN-\Lambda
_c\Lambda $ mixing. In view of
current studies of double-strangeness hypernuclei \cite{Myint,LY2}, one can say that
there exists very strong baryonic mixing in strange charmed
nuclei. To our knowledge, nobody studied the mixing of baryonic states in $C=+1$, $S=-1$ nuclei, which are possibly not pure $\Xi _c$ states, but rather superpositions of $\Xi _cN$ and $\Lambda
_c\Lambda $} states.

{\item $\Lambda_b$ nuclei were considered \cite{Kol,Star,Tsush,Tan} in similar lines. Differences between $\Lambda_c$ 
and $\Lambda_b$ nuclei originate evidently from the greater mass and zero electric charge of $\Lambda_b$.}
\end{enumerate}

\section{Features of the CSA; advantages and disadvantages}
The starting point of the CSA, as well as of the chiral perturbation theory, is the effective chiral lagrangian
written in terms of the chiral fields incorporated into the unitary matrix $U \in SU(2)$ in the original variant of
the model \cite{skyrme1,witten}, $U=cos\,f + isin\,f\; \vec\tau \vec n$, $n_z=cos\alpha,\,n_x=sin\alpha\,cos\beta,\,n_y=sin\alpha\sin\beta$,
where functions $f$ (the profile of the skyrmion), and angular functions $\alpha,;\beta$ in general case are
the functions of 3 coordinates $x,y,z$.
To get the states with flavor $S,\; c$ or $b$ we make extension of the basic $U\in SU(2)$ to $U\in SU(3)$
with $(u,d,s)$, $(u,d,c)$ or $((u,d,b)$.

It is convenient to write the lagrangian density of the model in terms of left (or right) chiral derivative
$$ l_\mu = \partial_\mu U U^\dagger = - U\partial_\mu U^\dagger \eqno (3.1) $$

$$ {\cal L} = -{F_\pi^2\over 16}l_\rho l_\rho +{1\over e^2} \left[l_\rho l_\tau\right]^2+
{F_\pi^2 m_\pi^2\over 16} Tr (U + U^\dagger - 2 )  \eqno (3.2) $$

Mass splittings $\delta M$ are due to the term in the lagrangian
$$ {\cal L}_M \simeq - \tilde m_D^2 \Gamma {s_\nu^2\over 2} \lambda_8, \eqno (3.3)$$
$s_\nu = sin (\nu)$, $\nu$ is the angle of rotation into "strange" direction, 
$\tilde m_K^2 = F_D^2m_D^2/F_\pi^2 -m_\pi^2$ includes the $SU(3)$-symmetry violation in
flavor decay constants, the quantity $\Gamma$, proportional to the sigma - term 
$$ \Gamma (\lambda)\simeq {F_\pi ^2\over 2} \int (1-c_f) \lambda^3d^3r. \eqno (3.4) $$
Numerically, for the baryon number $B=1$ configuration, $\Gamma \sim 6\, Gev^{-1} $ \footnote{The contribution of the chiral and flavor symmetry
breaking mass terms into the baryon mass equals to
$$\delta M^{SB} = {m_\pi^2\over 2} \Gamma + \tilde m_K^2 C_S \Gamma, $$
where the first part may be interpreted as sigma-term, $\Sigma = m_\pi^2\Gamma/2$. For $\Gamma= 6\,Gev^{-1}$
we get $\Sigma \simeq 57 MeV$.}
moments of inertia $\Theta_\pi \sim (5 - 6)Gev^{-1},\;\Theta_K \sim (2 - 3)Gev^{-1}$, 
see \cite{vkpent,kopshun} and references here. All moments of inertia $\Theta \sim N_c$.

The advantage of the CSA consists in the possibility to consider baryonic states with different flavors -
strange, charmed or beautiful - and with different atomic (baryon) numbers from unique point of view, using
one and the same set of the model parameters. The properties of the system are evaluated as a function
of external quantum numbers which characterize the system as a whole, whereas the hadronic content of the state 
plays a secondary role. This is in close correspondence with standard experimental situation where e.g.
in the missing mass experiments the spectrum of states is measured at fixed external quantum numbers
- strangeness or other flavor, isospin, etc. The so called deeply bound antikaon-nuclei states have been
considered from this point of view in \cite{koppot} not in condratiction with data (this is probably one of
most striking examples).

Remarkably that the moments of inertia of multiskyrmions carry information about their interactions. 
Probably, the first example how it works are the moments of inertia of the toroidal $B=2$ biskyrmion.
The orbital moment of inertia $\theta_J$ is greater than the isotopic moment of inertia $\theta_I$,
as a result, the quantized state with the isospin $I=0$ and spin $J=1$ (analogue of the deuteron) has smaller
energy than the state with $I=1,\;J=0$ (quasi-deuteron, or nucleon-nucleon scattering state), in qualitative 
agreement with experimental observation that deuteron is bound stronger \cite{kopax, bracar}.

The total binding energies of strangeness $S=-1$ hypernuclei have been estimated in \cite{kop-2} 
in qualitative agreement with data.
Much more successful was the description of the so called symmetry energy of nuclei with atomic numbers up to
$\sim 32$ and isospin up to $\sim 4 \,-\,5$ (i.e neutron excess up to $\sim 10$) \cite{ksm}. 
The variant of the model with the 6-th order term in chiral derivatives in the lagrangian density has been 
included, but flavors strangeness, charm or beauty have not been involved in this consideration.

Recently a variant of the model with the 6-th order stabilizing term in the lagrangian attracted much
attention \cite{marl1} - \cite{adam2}, and it has been noted that the binding energies of heavy nuclei
are described better in this variant than in the original variant with the Skyrme stabilizing term.

In view of this moderate success we can hope that further studies of baryonic states with different quantum
numbers in framework of the CSA, including states with unusual properties, may be of interest and useful.

Previously estimates of the flavor excitation energies were made mostly in perturbation theory,
i.e. the flavor excitation energy has been simply added to the skyrmion energy. This is not justified,
however, when the flavor excitation energy is large. Here we include this energy into simplified minimization
procedure which is made by means of the change of the soliton dimension (rescaling of the soliton).
This procedure takes into account the main degree of freedom of skyrmions given by the rational map anzatz \cite{hms} 
and leads to considerable decrease of the energy of states.
This modification of the skyrmion was made, in particular, by B.Schwesinger et al \cite{kss} to improve
the description of strange dibaryon configurations.

\section{Static properties of multiskyrmions}
In this section we present some static properties of multiskyrmions which are 
necessary to perform the procedure of the $SU(3)$ quantization and to obtain the spectrum of states with
definite quantum numbers.

The flavored moment of inertia equals within the rational map approximation for the original variant of the model
with the 4-th order in chiral derivatives term as the soliton stabilizer (the $SK4$ variant, 
we added the rescaling factor - some power of the parameter $\lambda $ to make evident the behaviour
under the rescaling procedure $r\to r\lambda$)

$$ \Theta_F^{SK4} = \lambda f_1 + \lambda^3 f_3^{(0)} {F_D^2 \over F_\pi^2} = 
\theta_F^{(0)} + \lambda^3 f_3^{(0)} \left({F_D^2 \over F_\pi^2} -1\right) \eqno (4.1) $$

with

$$ f_1 ={\pi\over 2 e^2} \int (1-c_F)\biggl(f'^2 +2B {s_f^2\over r^2}\biggr) r^2 dr; \qquad
 f_3^{(0)} ={\pi \over 2} F_\pi^2 \int (1- c_F) r^2 dr. \eqno (4.2)$$

Here we show explicitly the dependence of different parts of the inertia on the rescaling parameter $\lambda$.
In Table 1 we present numerical values for  $f_1$ and $f_3$.

There is simple connection between total moment of inertia in the $SK4$ variant of the model, the
$\theta_F^{SK4} $ and the sigma-term:
$$ \theta_F^{tot,SK4} = {F_D^2\over 4F_\pi^2}\Gamma + \theta_F^{SK4} = {F_D^2\over F_\pi^2}f_3^{(0)} + f_1. \eqno(4.3) $$

\begin{center}
\begin{tabular}{|l|l|l|l|l|l||l|l|l|l|l|}
\hline
$B$& $t_1(SK4)$&$f_1(SK4)$&$t_3(SK4)$&$f_3(SK4)$&$\Gamma(SK4)$&$t_1^*(SK4)$&$f_1^*(SK4)$&$t_3^*(SK4)$&$f_3^*(SK4)$&\\
\hline
$1 $ & $2.64$  & $0.85$    &$2.92$      &$1.20$       & $4.80$ & $6.67$ & $2.14$   &$6.13$ & $2.52$ & \\ \hline
$2 $ & $6.69$  & $1.84$    &$4.81$      &$2.34$       & $9.35$ & $13.57$ & $4.64$   &$10.73$& $5.22$ &\\ \hline
$3 $ & $8.14$  & $2.84$    &$6.26$      &$3.50$       & $14.0$ & $20.53$ & $7.18$   &$14.17$& $7.92$ & \\ \hline
$4 $ & $9.37$  & $3.77$    &$7.43$      &$4.50$       & $18.0$ & $26.35$ & $9.38$   &$16.55$& $10.02$ & \\ \hline
$5 $ & $14.67$ & $4.85$    &$8.83$     &$5.95$       & $23.8$ & $33.76$ & $12.1$    &$19.74$& $13.3$ &\\ \hline
$6 $ & $15.39$ & $5.85$    &$10.01$    &$7.25$       & $29.0$ & $40.27$ & $14.5$    &$22.33$& $16.18$  & \\ \hline
$7 $ & $18.19$ & $6.62$    &$10.71$    &$8.08$       & $32.3$ & $45.57$ & $16.8$    &$24.03$& $18.12$ &\\ \hline
$8 $ & $21.41$ & $7.68$    &$11.99$    &$9.72$       & $38.9$ & $52.96$ & $19.4$    &$26.94$& $21.85$  &\\ \hline
$9 $ & $24.28$ & $9.02$    &$13.52$    &$11.58$      & $46.3$ & $59.41$ & $21.8$    &$31.61$& $25.25 $   &\\ \hline
$10$ & $26.85 $& $10.0$    &$14.55$    &$13.0 $      & $52.0$ & $65.79$   & $24.4$    &$33.74$& $28.25$    &\\ \hline

$11$ & $29.54$ & $10.98$   &$15.66$    & $14.62 $    & $58.5$ & $72.26$ & $27.0$    &$33.74$&31.5 & \\ \hline
$12$ & $31.91$ & $11.98$   &$16.59$    & $16.02$     & $64.1$ & $78.28$ & $29.3$    &$35.72$&34.5 & \\ \hline
$13$ & $34.54$ & $12.95$   &$17.56$    & $17.55$     & $70.2$ & $84.23$ & $31.8$    &$37.77$&37.75 &\\ \hline
\end{tabular}
\end{center}
{\bf Table 1.} 
The Skyrme term contribution to the isotopic moment of inertia of multiskyrmions $\Theta_I({\rm SK4})$, 
the "flavor" inertia $\Theta_F({\rm SK4})$ 
and the sigma term $\Gamma$ in the SK4 variant of                                                                                                                    2
the model with $e=4.12$ and in the rescaled variant with $e=3$ (columns 7 --- 10),  in ${\rm GeV}^{-1}$. \\

Similarly, the isotopic moment of inertia $\theta_I$ within the rational map aproximation can be written as

$$ \Theta_I^{SK4} = \lambda \,t_1 + \lambda^3 \,t_3  \eqno (4.4) $$

with
$$ t_1 ={4\pi\over 3} \int  {2\,s_F^2 \over e^2}\left(F'^2 +B{s_F^2\over r^2}\right) r^2 dr, \qquad t_3 =
{2\pi\over 3} F_\pi^2 \int s_F^2  r^2 dr. \eqno (4.5) $$

In the $SK6$ variant of the model the skyrmion stabilization takes place due to the 6-th order term (in chiral derivatives) in the lagrangian density,
which is proportional to the baryon number density squared \cite{kjp}.

In Table 2 we present numerical values for the contributions to the moments of inertia, which scale differently under change of the soliton scale;

$$ \Theta_F (SK6) = f_3 \lambda^3 + f_6/\lambda, \eqno (4.6) $$

$$\Theta_I (SK6) = t_3 \lambda^3 + t_6/\lambda, \eqno(4.7) $$

$$  t_6 = {1\over 8} \int (1-c_f)2 c_6s_f^2\biggl(2Bf'^2 + {\cal I}{s_f^2\over r^2}\biggr) dr. \eqno (4.8) $$

The relation takes place in the $SK6$-model, analogous to previous one:
$$ \theta_F^{tot,SK6} = {F_D^2\over 4F_\pi^2}\Gamma + \theta_F^{SK6}. \eqno (4.9) $$

$$\theta_F (\lambda) ={\pi\over 2} \int (1-c_f)\Biggl[\lambda^3 F_D^2 + {\lambda \over e^2}\biggl(f'^2 +2B {s_f^2\over r^2}\biggr)+
2 {c_6\over \lambda} {s_f^2\over r^2}\biggl(2Bf'^2 +{\cal I} {s_f^2\over \,r^2}\biggr)\Biggr]r^2 dr \eqno (4.10) $$

To calculate the part of the isotopic inertia proportional to $F_\pi^2$ we used
analytical approach developed in \cite{kjp}.
 Under scale transformation we have $\theta_F^{SK6} \sim \lambda^{-1} $.

\begin{center}
\begin{tabular}{|l|l|l|l|l|l|l|}
\hline
$B$& $\Theta_I({\rm SK6})$&$\Theta_F({\rm SK6})$&$\Gamma({\rm SK6})$&$\Theta_I({\rm SK6*})$&
$\Theta_F({\rm SK6*})^*$&$\Gamma({\rm SK6*})^*$\\
\hline
$1 $ & $5.13$ & $0.76$ & $6.08$ & $14.2$ & $2.38$ & $15.3$ \\ \hline
$2 $ & $9.26$ & $1.44$ & $14.0$ & $25.7$ & $4.62$ & $35.9$ \\ \hline
$3 $ & $12.7$ & $2.18$ & $20.7$ & $35.5$ & $6.92$ & $53.9$ \\ \hline
$4 $ & $15.2$ & $2.80$ & $24.5$ & $43.2$ & $8.85$ & $64.6$ \\ \hline
$5 $ & $18.7$ & $3.60$ & $32.8$ & $52.9$ & $11.35$ & $86.2$ \\ \hline
$6 $ & $21.7$ & $4.28$ & $39.3$ & $61.4$ & $13.65$ & $103 $ \\ \hline
$7 $ & $23.9$ & $4.88$ & $42.5$ & $68.0$ & $15.3$ & $112 $ \\ \hline
$8 $ & $27.2$ & $5.60$ & $51.6$ & $77.3$ & $17.9$ & $135 $ \\ \hline
$9 $ & $30.2$ & $6.32$ & $59.1$ & $85.7$ & $20.4$ & $154 $ \\ \hline
$10$ & $32.9$ & $7.05$ & $65.8$ & $93.5$ & $22.6$ & $171 $ \\ \hline

$11$ & $35.8$ & $7.70$ & $73.6$ & $102 $ & $24.8$ & $191 $ \\ \hline
$12$ & $38.4$ & $8.32$ & $79.9$ & $109 $ & $27.0$ & $207 $ \\ \hline
$13$ & $41.2$ & $9.02$ & $87.1$ & $117 $ & $29.2$ & $225 $ \\ \hline
\end{tabular}
\end{center}

{\bf Table 2.} Same as in Table 1,  for the SK6 variant of the model.
 $e' =4.11$ and the rescaled variant of the model, $e' =2.84$ . \\

Here $\theta_F^{SK6}$ scales like $1/\lambda$.

\section{Flavor excitation energies and the total binding energies estimates}
The total binding energies are estimated using the double subtraction procedure.

We shall use the following mass formula for the quantized state which allowed to
estimate the binding energies of hypernuclei in \cite{kop-2}
$$M(B,F,I,J) = M_{cl}+ |F|\omega_{F,B} +{1\over 2\theta_{F,B}}
\left[c_FI_r(I_r+1)+(1-c_F)I(I+1)+(\bar c_F-c_F)I_F(I_F+1)\right]+ $$
$$ +{J(J+1)\over 2\theta_J}, \eqno (5.1)$$
where index B has been omitted for the sake of brevity.
For $|F|=1$ we have in present paper $I_F=1/2$.

\begin{center}
\begin{tabular}{|l|l|l|l|l|l|l|l|l|}
\hline
$B$& $\omega_c({\rm SK4})$&$\omega_b({\rm SK4})$&$\omega_C({\rm SK4*})$&
$\omega_b({\rm SK4*})$&$\omega_c({\rm SK6})$&$\omega_b({\rm SK6})$
&$\omega_c({\rm SK6*})$&$\omega_b({\rm SK6*})$\\
\hline
$1 $ & $1.54$ & $4.80$ & $1.55$ & $4.77$ & $1.61$ & $4.93$ & $1.62$  &$4.89$ \\ \hline
$2 $ & $1.52$ & $4.77$ & $1.54$ & $4.75$ & $1.64$ & $4.98$ & $1.66$ & $4.95$ \\ \hline
$3 $ & $1.51$ & $4.76$ & $1.54$ & $4.74$ & $1.64$ & $4.98$  &$1.66$ & $4.95$ \\ \hline
$4 $ & $1.50$ & $4.74$ & $1.52$ & $4.72$ & $1.62$ & $4.92$ &$1.64$  &$4.93$ \\ \hline
$5 $ & $1.51$ & $4.75$ & $1.53$ & $4.74$ & $1.63$ & $4.96$ &$1.65$  &$4.94$ \\ \hline
$6 $ & $1.51$ & $4.76$ & $1.54$ & $4.74$ & $1.634$ &$4.96 $&$1.65$  &$4.94$ \\ \hline
$7 $ & $1.50$ & $4.74$ & $1.53$ & $4.73$ & $1.623$ &$4.95 $&$1.64$  &$4.93$ \\ \hline
$8 $ & $1.51$ & $4.76$ & $1.54$ & $4.75$ & $1.63$ & $4.96 $&$1.65$  &$4.94$ \\ \hline
$9 $ & $1.52$ & $4.77$ & $1.54$ & $4.76$ & $1.63$ & $4.97 $&$1.65$  &$4.94$ \\ \hline
$10$ & $1.52$ & $4.78$ & $1.558$ &$4.76$ & $1.63$ & $4.97 $&$1.65$  &$4.94$ \\ \hline

$11$ & $1.53$ & $4.79$ & $1.55$ & $4.77$ & $1.63$ & $4.97$ &$1.65$& $4.95$\\ \hline
$12$ & $1.53$ & $4.79$ & $1.55$ & $4.77$ & $1.63$ & $4.97 $&$1.65$ & $4.95$\\ \hline
$13$ & $1.53$ & $4.79$ & $1.55$ & $4.77 $ &$1.63$ & $4.98 $&$1.65$  & $4.95$\\ \hline
\end{tabular}
\end{center}

{\bf Table 3.} Flavor excitation energies for charm and beauty, for the
$SK4$ and $SK6$ variants of the model. Calculations are made using the ratios
$F_D/F_\pi = 1.576$ and $F_B/F\pi \simeq 1.44$ according to \cite{na}.\\

The flavor excitation energy is
$$\omega_{F,B} = {3B\over 8\theta_{F,B}} (\mu_{F,B} -1 ) \eqno (5.2) $$
with
$$\mu_{F,B} = \left[1+{16 \bar m_D^2\Gamma_B\theta_{F,B}\over 9B^2}\right]^{1/2}$$
   At large enough $m_D$ the expansion can be made
$$\mu_{F,B} \simeq {4\bar m_D(\Gamma_B \theta_{F,B})^{1/2}\over 3B}+
{3B\over 8\bar m_D\Gamma_B \theta_{F,B}}, \eqno (5.3)$$
therefore 
$$\omega_{F,B} \simeq {1\over 2}\bar m_D
\left({\Gamma_B\over \theta_{F,B}}\right)^{1/2}
 -{3B\over 8\theta_{F,B}}.  \eqno (5.4) $$

\section{Hyperfine splitting correction to the energy of the state}
The correction to the energy of states which is formally of the $1/N_c$ order
in the number of flavors has been obtained previously in \cite{kw,ks}
$$\Delta E_{1/N_c} ={1\over 2\theta_I}\left[c_FI_r(I_r+1) +(1-c_F) I(I+1) +
(\bar c_F-c_F) I_F(I_F+1)\right] \eqno (6.1) $$
where index $B$ is omitted for the sake of brevity, $I$ is the isospin of the state,
$I_F$ is the isospin carried by flavored meson $(K,\,D,$ or $B-$meson, for unit 
flavor $I_F=1/2$,  $I_r$ can be 
interpreted as "right" isospin, or isospin of basic non-flavored configuration.
The hyperfine splitting constants
$$c_F= 1 -  {\theta_I(\mu_F - 1) \over 2\theta_F\mu_F}, \quad  
\bar c_F= 1 -  {\theta_I(\mu_F - 1) \over \theta_F\mu_F^2},  \eqno (6.2) $$

This correction is considered usually as small one, but in some cases it can be
included into the minimization procedure, e.g. when isospin $I$ is large.

In our previous calculations we used also the following expression for 
the difference of energies between the state with flavor $|F|$ isospin $I$ and 
the state with zero flavor $F$, isospi $I_r$ which belongs to same $SU(3)$
multiplet $(p,q)$:

$$\Delta E(B,F) = |F|\omega_F +{\mu_F -1\over 4\mu_F\theta_F}[I(I+1)-I_r(I_r+1)]
+{(\mu_F-1)(\mu_F-2)\over 4\mu_F^2\theta_F} I_F(I_F+1) \eqno (6.3) $$
and we considered the flavor excitation energy as small perturbation which
makes no influence on the skyrmion itself. Such approach may be justified only
for strangeness (but should be checked in this case as well), not for charm or
beauty quantum numbers where this energy is large.

\section{Rescaling procedure}
The flavor excitation energy and hyperfine splitting correction have been 
considered previously as small corrections to the energy of the state.
Such an approach can be, however, not justified when the flavor excitation is
not so small, as for heavy flavors, charm and beauty.

In these cases it is reasonable to include into consideration the
overall scale of the soliton and to perform further minimization of
the energy as a function of this scaling parameter.

Let us consider several examples of interest.

\subsection{Even $B$-number, $|F| = 1$}
For the case of even baryon number, not very large, the ground state of the nucleus
has zero isospin, and it belongs to the $SU(3)$ multiplet $(p,q) = (0,\,3B/2)$.
So, in the above formula we have
$$I=0,\quad I_r = 0,\quad I_F = 0. \eqno (7.1) $$
The energy of this state coincides with the classical mass of the skyrmion:
$$ E(B,0,0,0,0) = M_B^{cl}. \eqno (7.2) $$

For the flavored state with $|F|= 1$ we should take
$$I_r=0,\quad I=I_F= {1\over 2} \eqno (7.3) $$
and the energy of this state is
$$E(B,|F|=1,I_F=1/2, I_r=0, I=1/2) = M_B^{cl} + \omega_F +{3(\mu_F -1)^2 \over 4\theta_F \mu_F^2}. \eqno (7.4) $$

As a next step we should write this energy as function of rescaling parameter $x$ and
to find minimal energy $E^{min}(x_{min}$.

The next step is to estimate the change of the binding energy of state, when substitution
of the nucleon by the $\Lambda-$hyperon was made.

\subsection{Odd $B$-numbers, $|F|=1$}

In this case for the ground state we have $I_F=0,\; I=I_r=1/2$.
The energy of these states
$$ E(B, 0,0,1/2,1/2) = M_B^{cl} + {3\over 8\theta_I}. \eqno (7.5)  $$

For flavored states with $|F|=1,\, I_F=1/2, I_r=1/2,\,I=0$ its energy is
$$E(B,1, 1/2, 1/2, 0) = M_B^{cl} +\omega_F +{3\over 8\theta_I} - {3(\mu_F-1) \over 8\theta_F \mu_F^2}  \eqno (7.6) $$
and these energies should be minimized as functions of rescaling parameters.

The $B=1$ case should be considered in similar way.
The nucleon mass, $I=I_r=1/2, \;  I_F=0$:

$$M_N = M_1^{cl} + {3\over 8\theta_{I,1}}  \eqno (7.7) $$

The $\Lambda-$hyperon mass

$$M_\Lambda = M_1^{cl} + \omega_{F,1} + {3\bar c_{F,1}\over 8\theta_{I,1}}  = 
M_1^{cl} + \omega_{F,1} + {3\over 8\theta_{I,1}} - {3(\mu_{F,1}-1)\over 8\mu_{F,1}\theta_{F,1}}. \eqno (7.8) $$

Both $M_N$ and $M_\Lambda$ should be minimized separately with own scaling parameter $x$.

We do not pretend to calculate the binding energies, but we can estimate the changes in binding energies
of baryonic systems with flavor (hypernuclei) in comparison with nonflavored baryonic system:
$$ \Delta_\epsilon (B,F) = - E_{B,F} + M_B + M_\Lambda - M_N . \eqno (7.9) $$

The numerical results are presented here.

\section{Conclusions and prospects}
We have estimated the total binding energies of baryonic states (hypernuclei)
with quantum numbers charm or beauty, and some neutron excess, or high isotopic spin.

The rescaling procedure is important for several values of baryon number,
and with increasing B-numbers it becomes less important. 


We thank Yura Ivanov for important help in numerical computations.

\section{Appendix 1. Analytical treatment of multiskyrmions properties in  the rational map approximation}
The rational map approximation for multiskyrmions, proposed in \cite{hms}, allows to get analytical expressions describing
characteristics of multiskyrmions (masses, moments of inertia, sigma term) \cite{kjp}, 
valid with an acuracy of several percents.

Starting point of the anlytical treatment is parametrization of the multiskyrmion profile function in the form
$$\phi = cos\,F = \frac{(r/r_0)^b - 1}{(r/r_0))^b + 1}, \eqno(A1.1) $$
which has correct boundary conditions $cos\,f(0) = -1$, $cos\,f(\infty) = 1$, and the constants $r_0(B)$ - dimension of
the skyrmion, $ b(B)$ - the effective power, depending on the baryon number of multiskyrmion, can be found by the
static mass minimization procedure \cite{kjp}. 
The integrals over 3-dimensional space which appear in calculation of the skyrmion mass, as well as other skyrmions 
properties are the Euler-type integrals which can be evaluated in general enough form

$$ \int_0^\infty \frac{(r/r_0)^c dr}{\beta+(r/r_0)^b} = \beta^{(1+c)/b-1}\frac{\pi\,r_0}{b\, sin[\pi(1+c)/b]}. \eqno(A1.2) $$
The values of $r_0^{min}$ and $b^{min}$ were found for the $SU(2)$ model
with the Skyrme term as the skyrmion stabilizer to be \cite{kjp}
$$ r_0^{min} \simeq [2 \sqrt{\cal I}/3]^{1/2}, \qquad b_0^{min}\simeq 2 {\cal I}^{1/4}. \eqno(A1.3)  $$

The mass term (or sigma-term) is proportional to
$$ \Sigma = \int (1 - cos\,F) d^3 r = {8\pi^2 r_0^3\over b\,sin (3\pi/b)}. \eqno (A1.4)  $$
The contribution to the flavor moment of inertia due to the second order (kinetic) term in the lagrangian equals to
$$ \Theta_F^{(2)}  =  \int sin^2F d^3r  = {48 \pi^2 r_0^3 \over b^2 sin (3\pi/b)}, \eqno (A1.5) $$
see Eq. (42) of \cite{kjp}.
Using analytical approach, we obtain simple relations between different quantities of interest.

\section{Appendix 2. The classical mass of the skyrmion, and corrections}
The classical mass of the soliton can be written in form (original $SK4$ variant of the model)
$$ M^{cl} = \lambda m_1  + m_2/ \lambda  + \lambda^3\,m_3. \eqno (A2.1)  $$

The quantum correction depends on moments of inertia which can be expanded in similar way.
The flavor moment of inertia

$$\theta_F =\lambda\, f_1 + \lambda^3\,f_3 ,  \eqno(A2.2)  $$

and isotopic moment of inertia

$$ \theta_I = t = \lambda\,t_1 + \lambda^3 t_3 .\eqno (A2.3) $$

These expressions should be substituted to the expression for the energy of the state, and the 
minimum should be found.

For the ground states with odd baryon numbers we should find the minimum of the
energy $(F=o, I_F=0, I=I_r = 1/2)$as function of scale parameter $\lambda$;

$$ E_{gr}(odd) = M_{cl} +  {3\over 8 \Theta_I} = \lambda m_1 + m_2/ \lambda  +\lambda^3\,m_3 +  {3\over \lambda t_1 + \lambda^3 t_3 } \eqno (A2.4) $$

The energy of the state with unit flavor $(|F|=1, I_F=1/2, I=0)$ can be eathearly written.  


\end{document}